\documentclass[11pt,twoside]{article}

\usepackage{prcsa2008}
\usepackage{epsf}
\usepackage{psfig}
\usepackage{lscape}

\begin{document}
\title{Spectroscopic Observations of WZ Sge-type Dwarf Novae, GW Lib and V455 And in Superoutburst}   %%% Fill in title
\author{D.~Nogami\altaffilmark{1}, K.~Hiroi\altaffilmark{2},
Y.~Suzuki\altaffilmark{2}, Y.~Moritani\altaffilmark{2},
Y.~Soejima\altaffilmark{2}, A.~Imada\altaffilmark{3},
O.~Hashimoto\altaffilmark{4}, K.~Kinugasa\altaffilmark{4},
S.~Honda\altaffilmark{4}, K.~Ayani\altaffilmark{5},
S.~Narusawa\altaffilmark{6}, H.~Naito\altaffilmark{6},
M.~Sakamoto\altaffilmark{6}, T.~Iijima\altaffilmark{7},
M.~Fujii\altaffilmark{8}, N.~Narita\altaffilmark{9}}   %%% Fill in author names
%\affil{}    %%% Fill in author affiliations
\altaffiltext{1}{\it Hida Observatory, Kyoto University, Kamitakara, Gifu
506-1314, Japan}
\altaffiltext{2}{\it Dept. of Astronomy, Fac. of Science, Kyoto University,
Sakyo-ku, Kyoto 606-8502, Japan}
\altaffiltext{3}{\it Dept. of Physics, Fac. of Science, Kagoshima University, 1-21-35 Korimoto, Kagoshima 890-0065, Japan}
\altaffiltext{4}{\it Gunma Astronomical Observatory, 6860-86 Nakayama Takayama, Agatsuma, Gunma 377-0702, Japan}
\altaffiltext{5}{\it Bisei Astronomical Observatory, 1723-70 Ohkura, Bisei, Okayama 714-1411, Japan}
\altaffiltext{6}{\it Nishi-Harima Astronomical Observatory, Sayo-cho, Hyogo 679-5313, Japan}
\altaffiltext{7}{\it Astronomical Observatory of Padova, Asiago Section, Osservatorio Astrofisico, 36012 Asiago (Vi), Italy}
\altaffiltext{8}{\it Fujii-Bisei Observatory, 4500 Kurosaki, Tamashima, Okayama 713-8126, Japan}
\altaffiltext{9}{\it National Astronomical Observatory of Japan, 2-21-1 Osawa,
Mitaka, Tokyo 181-8588, Japan}

\begin{abstract} %%% Abstract to run on from here.
We carried out intensive spectroscopic observations of two WZ Sge-type
dwarf novae, GW Lib, and V455 And during their superoutbursts in 2007,
at 6 observatories.  The observations covered the whole of both
superoutbursts from the very maximum to the fading tail.  We found
evidence of the winds having a speed of $\sim$1000 km s$^{-1}$ which
blew in GW Lib during the rising phase.  The evolution of the hydrogen,
helium, and carbon lines suggests flaring of the accretion disk and
emergence of the temperature inversion layer on the disk.
\end{abstract}

%%% MAIN BODY OF TEXT GOES HERE. CONSULT "INSTRUCTIONS FOR AUTHORS USING
%%% LATEX2E MARKUP", SECTIONS 2.3-2.6 FOR HELP WITH EQUATIONS, FIGURES,
%%% AND TABLES.

\section{Introduction}

Dwarf novae are a class of cataclysmic variable stars, which show outbursts
by the disk instabilities, and SU UMa-type dwarf novae are a subclass giving
rise to two types of outburst: normal outbursts and superoutbursts
\citep[for a review][]{Osa96}.  There is a small group of SU UMa
stars, being called WZ Sge-type dwarf novae.  Their peculiar feature is
summarized as: the large outburst amplitude over 6 mag, the long recurrence
cycle of the superoutburst (several years or more), and no (or few) normal
outburst.  In addition, they show curious behavior in many points, such as 
``early'' superhumps, a variety of rebrightening light curves after the main
superoutburst, long fading tail, and so on \citep[see][and references
therein]{Uem08}.

The driving mechanisms working in WZ Sge stars are still in debate
\citep[see][]{nog07}, while the basic properties of normal SU UMa-type dwarf
novae are generally explained by the thermal-tidal disk instability theory.
For further investigation, we have prepared for coordinated spectroscopic
observations during forthcoming superoutbursts, following a success of
intensive spectroscopy during the 2001 superoutburst in WZ Sge itself
\citep{bab02,nog04}.

Under this circumstances, superoutbursts of two WZ Sge-type dwarf novae,
GW Lib, and V455 And (=HS2331+3905) were discovered in the very early phase
in 2007 by eager contributors to VSNET \citep[about VSNET, see][]{kat04}.
We succeeded in soon starting coordinated spectroscopic campaigns before
or around the maximum in both stars, and here review the preliminary results
and implications.  The details will be published in our forthcoming papers.

\section{GW Librae}

GW Lib was discovered during an outburst at 9 mag as a novalike object
in 1983 \citep{maz83}, and is the first cataclysmic variable where
white dwarf pulsations with periods of hundreds-thousands of seconds
were found \citep{wou02,van04}.  \citet{tho02} measured its orbital
period to be 0.05332(2) d, and estimated the inclination to be $\sim$11
deg.

After dormancy of 23 years since the discovery, an outburst was caught
at 13.8 mag at 2007 April 12.494 (UT) (R. Stubbings, vsnet-alert 9279).
Our spectroscopic observations started at April 12.66 (UT).

\begin{figure}[!ht]
\plottwo{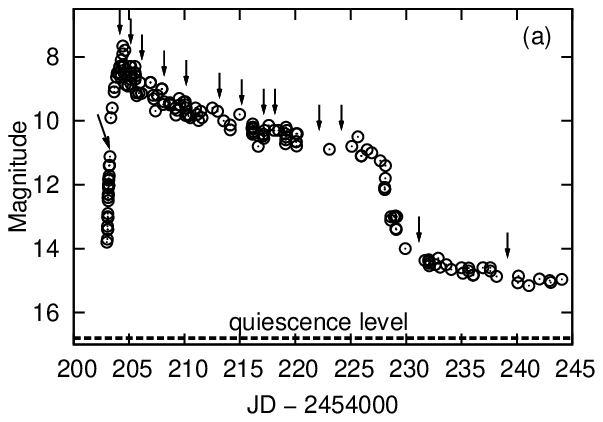}{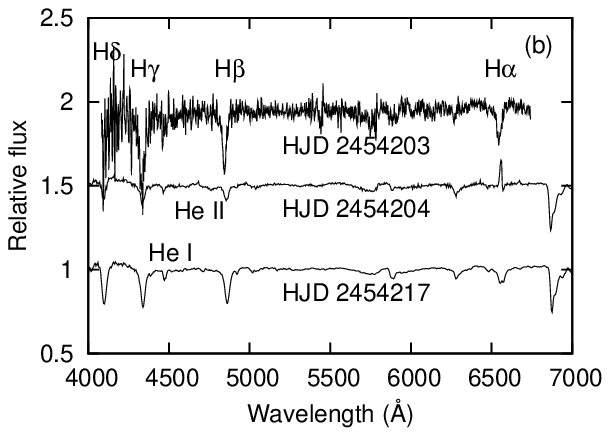}
\caption{(a) Whole light curve of the 2007 superoutburst in GW Lib generated
with the data reported to VSNET.  The down arrows indicate the date when the
spectra were obtained.  Our intensive campaign covers the rising phase to
the long fading tail.  (b) Representative normalized spectra before the maximum,
at the maximum, and in the plateau phase from upper to lower.  The spectrum
before the maximum shows only Balmer lines in absorption which are blue-shifted
by $\sim$1000 km s$^{-1}$.  At the maximum, we can see the narrow H$\alpha$
emission line, and other Balmer absorption lines with an emission core.  The
He II 4686 emission line associated by the Bowen blend C III/N III is also
present.  The emission component of the Balmer lines gradually decayed during
the plateau phase.}
\end{figure}

The light curve of the superoutburst cut on the way of the long fading tail,
and the observation date are displayed in figure 1a.  Figure 1b exhibits
the representative spectra, normalized to a unity continuum value, before the
maximum, at the maximum, and in the plateau phase.  All the spectra had a
very blue continuum before the normalization, as is often seen in dwarf novae
in outburst.

Before the maximum, the spectrum had only Balmer absorption lines.  The central
wavelength of the Balmer lines are blue-shifted by $\sim$1,000 km s$^{-1}$, and
the full width at the half maximum is about 3,000 km s$^{-1}$.  Taking into
account that GW Lib is a nearly pole-on system, these results indicate that
optically thick winds blew just after the onset of the outburst.

By the maximum, H$\alpha$ turned to be a singly-peaked narrow emission line
(FWZI $<$ 1000 km s$^{-1}$).  There also appeared emission lines of highly
excited species, He II 4686, C III/N III, and possibly C IV/N IV around 5800
\AA, and a hint of absorption lines of Na I D.  These lines became weaker
during the plateau phase.  These imply that a temperature inversion layer
emerged on the accretion disk after switching off the winds, and this
layer gradually faded away.  Short-term variabilities in the line profile
were not detected in our time-resolved low-resolution spectra.

We can clearly see only Balmer emission lines in the spectra in the fading
tail.  They mimic those in quiescence, but the equivalent widths are much
smaller.

\section{V455 Andromedae}

V455 And was originally found as a CV candidate in the Hamburg Quasar Survey
\citep{hag95}.  Thorough observations by \citet{ara05} revealed its orbital
period of 81.08 min by using eclipses, 5-6 min variabilities resulting from
the WD pulsations, 1.12 min coherent oscillations attributable to the WD spins.
They also detected radial velocity modulations of Balmer and helium lines with
a period of $\sim$3.5 h.  Based on this report, this star has attracted
an attention as a candidate of WZ Sge stars.

First outburst of this star was detected in the very early phase at 2007
September 4.775 (UT) (H. Maehara, vsnet-alert 9530).  We then started
spectroscopy from September 5.38 (UT).

\begin{figure}[!ht]
\plotone{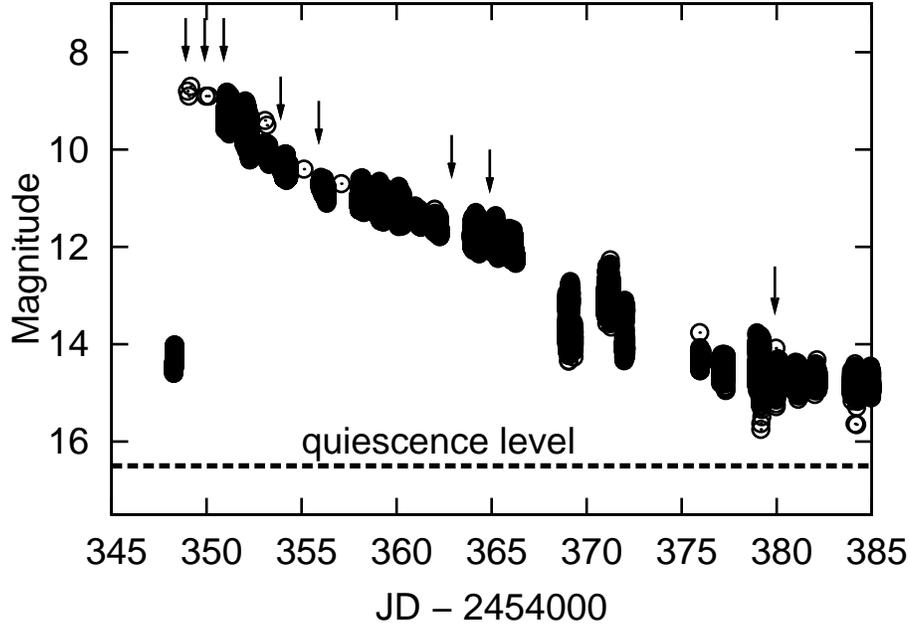}
\caption{Whole light curve of the 2007 superoutburst in V455 And generated
with the data reported to VSNET.  The down arrows indicate the date when the
spectra were obtained.  We took several more spectra during the long fading
tail which is not displayed in this figure.}
\end{figure}

The superoutburst light curve and the time when we obtained spectra are shown
in figure 2.  The light curve is cut on the way of the fading tail for a visual
purpose, but we obtained several spectra during the period not shown in figure
2.  The details of the photometric observations are reported by \citet{mae08}.

\begin{figure}[!ht]
\plottwo{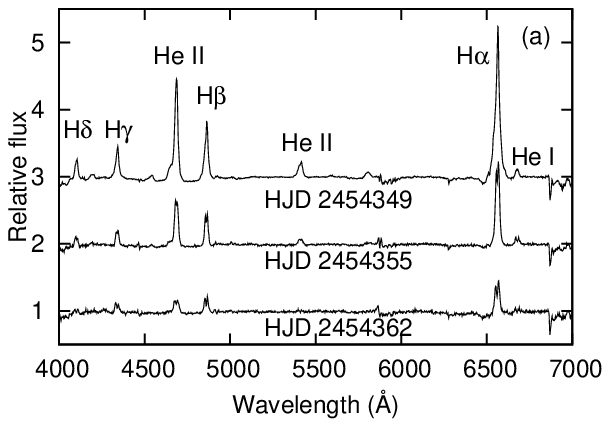}{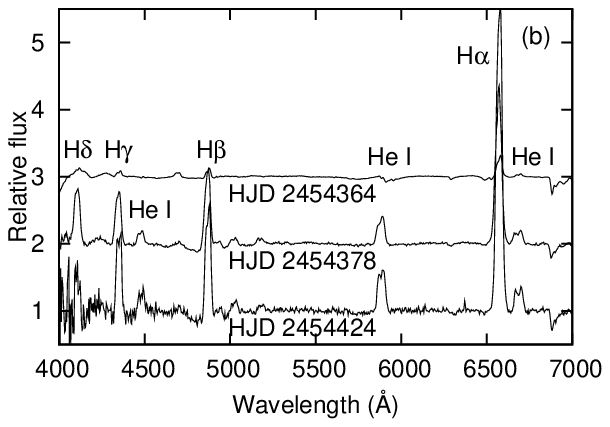}
\caption{Long-term spectral evolution indicated by representative spectra.
HJD 2454349 is around the maximum.  HJD 2454355, 2454362, and 254364 are in
the plateau phase.  HJD 2454378, and 2454424 are in the fading tail.  The
singly-peaked strong Balmer and He II lines at the maximum became weaker, as
the superoutburst proceeded.  During the fading tail, the doubly peaked
Balmer and He I emission lines grew again, as V455 And approached its
quiescence state.}
\end{figure}

The spectral line evolution is exhibited by the representative spectra in
figure 3.  The top spectrum in figure 3a was taken just at the maximum.
There exist strong emission lines of the Balmer series and He II.  We can
also see the Bowen blend C III/N III, and C IV/N IV, and He I in emission.
Note that all these emission lines have singly peaked shape, while the
spectrum in quiescence possesses doubly peaked emission lines \citep{ara05}.
As time passed, these emission lines became weaker during the plateau phase,
and doubly peaked shapes emerged.  These line-profile variabilities imply
the following scenario: 1) the accretion disk flared up by the outburst
maximum, and the singly peaked shape we observed is formed by significant
contributions of the temperature inversion layer on the disk edge, and 2)
the flaring disk gradually settled down during the plateau phase, and the
inner part of the accretion disk got observable, which resulted in the
doubly peaked emission lines.

During the long fading tail (HJD 2454378, and 2454424), the line spectra
are very similar to that in quiescence.  The equivalent width of the
Balmer and He I emission lines, however, grew up, as the the system
gradually faded.

\section{Disk Evolution}

We here summarize the plausible disk evolution during the superoutburst
in WZ Sge stars which are suggested by the intensive spectroscopic
observations of the nearly pole-on system, GW Lib, and the nearly edge-on
system, V455 And during the 2007 superoutbursts.

The winds with a speed of 1,000 km s$^{-1}$ blow just after the onset of
the superoutburst, and they drop by the maximum.  At the same time, the
disk flares up, and the temperature inversion layer is formed on the
accretion disk.  This flaring disk, and this layer gradually settle
down during the plateau phase.  After the rapid decline from the plateau
phase, the emissivity at the continuum gradually returns to the value
in quiescence, as that in the Balmer and He I lines do it more gradually.

More detailed analyses and quantitative discussion will be published in
our forthcoming papers.

\acknowledgements  The authors are grateful to the observers for reporting
their precious data to VSNET.  This work is partly supported by a grant-in-aid
from the Ministry of Education, Culture, Sports, Science, and Technology
(No. 17204012, 17740105).

\end{document}